\documentstyle[12pt,aaspp4]{article}
\begin{document}

\title{An Atlas of Warm AGN and Starbursts from the IRAS Deep 
Fields\footnote{
Based on observations from the European Southern Observatory, La Silla, Chile;
Kitt Peak National Observatory, National Optical Astronomy Observatories,
operated by the Asociation of Universities for Research in Astronomy, Inc., 
under cooperative agreement with the National Science Foundation; 
the Isaac Newton and William Herschel telescopes, operated
by the Royal Greenwich Observatory on behalf of the SERC at the Spanish
Observatorio del Roque de los Muchachos, and Lowell Observatory.}}

\author{William C. Keel, Bryan K. Irby\altaffilmark{2}, and Alana May}
\affil{Department of Physics and Astronomy, University of Alabama, Box 870324,
Tuscaloosa, AL 35487}

\altaffiltext{2}{Current address: L-3 Communications Government Services, Inc.,
Laboratory for High Energy Astrophysics,
NASA/Goddard Space Flight Ctr. Code 664, Greenbelt MD 20771.}

\author{George K. Miley}
\affil{Sterrewacht Leiden, Postbus 9513, 2300 RA Leiden, The Netherlands}

\author{Daniel Golombek}
\affil{Space Telescope Science Institute, 4700 San Martin Dr.,
Baltimore, MD 21218}

\author{M.H.K. de Grijp}
\affil{Marienpoelstraat 37,
2334 CX Leiden,
The Netherlands}

\author{Jack F. Gallimore}
\affil{Department of Physics, Bucknell University,
Lewisburg, PA 17837}

\begin{abstract}
We present a set of 180 AGN candidates based on color selection from the
IRAS slow-scan deep observations, with color criteria broadened from
the initial point-Source Catalog samples so as to include similar objects
with redshifts up to $z=1$ and allowing for two-band detections.
Spectroscopic identifications have been obtained for 80 (44\%);
some additional identifications are secure based on radio detections or
optical morphology, although yet unobserved spectroscopically. These
spectroscopic identifications 
include 13 type 1 Seyfert galaxies, 17 type 2 Seyferts, 29 starburst galaxies,
7 LINER systems, and 13 emission-line galaxies so heavily reddened as to 
remain of ambiguous classification. The optical magnitudes range from 
$R=12.0-20.5$; the counts suggest that incompleteness is important 
fainter than $R=15.5$. Redshifts extend to $z=0.51$, with a significant part
of the sample at $z>0.2$. Even with the relaxed color criteria, this
sample includes slightly more AGN than star-forming systems among those
where the spectra contain enough diagnostic feature to make the
distinction. The active nuclei include several broad-line
objects with strong Fe II emission, and composite objects with 
the absorption-line signatures of fading starbursts. These AGN with warm far-IR 
colors have little overlap with the ``red AGN" identified with 2MASS; only
a single Sy 1 was detected by 2MASS with $J-K > 2$. Some
reliable IRAS detections have either very faint optical counterparts or
only absorption-line galaxies, potentially being deeply obscured AGN. 
The IRAS
detections include a newly identified symbiotic star, and several
possible examples of the ``Vega phenomenon", including dwarfs
as cool as type K. Appendices detail these candidate stars, and
the optical-identification content of a particularly deep set of
high-latitude IRAS scans (probing the limits of optical identification
from IRAS data alone).
\end{abstract}


\section{Introduction}

Early assessment of galaxies in the IRAS Point-Source Catalog
demonstrated that many active galactic nuclei (AGN) have
very distinct colors among the IRAS bands, especially 25-60$\mu$
and 60-100$\mu$ (de Grijp et al 1985, 1987, 1992; Low et al. 1988). This 
meant that large samples of AGN could
be winnowed very effectively from the IRAS catalogs, to form
an all-sky sample which would prove to consist mostly of
Seyfert galaxies. Such samples, selected on infrared properties,
are important in studying the importance of obscuration in
Seyfert galaxies, and in probing possible low-redshift evolution
of IR-selected galaxies. These samples are dominated by
AGN which are luminous in comparison to IR emission from the
surrounding galaxy, since the IRAS colors refer to the entire
galaxy, combining the AGN and surrounding galaxy at these distances. 

In addition to the all-sky
survey, IRAS also observed $\approx 10^3$ fields in a slow-scan
mode to deeper flux levels, in the ``additional observations" or AOs.
These were usually targeted around some object of particular interest,
but their typical $0.5 \times 1.5^\circ$ extent meant that a substantial
region was also observed serendipitously. We report here a study
of objects selected from these IRAS AO data, using color criteria
based on the results of the earlier studies but widening the color
ranges to include typical ``warm" AGN at redshifts as high
as $z=1$. These fields go 3--7 times deeper than the Point-Source
Catalog (depending on ecliptic latitude and number of overlapping
AO observations of each field). Such a deep sample could probe intrinsically
rare objects, and span enough redshift range to detect strong
evolution. Since
the IR detections pertain to the whole galaxies, we expected that
much of the senstivity gain would result in finding similar
objects to greater distance, rather than going deeper into the
local AGN luminosity as often happened with X-ray surveys.
An initial report on the data processing and first AGN identifications
was presented by Keel et al. (1988).

The slower-scan AO observing mode resulted in less precise positions
than the all-sky survey, which combined with the fainter flux
limits to make optical identifications more difficult. In many cases,
slitless spectroscopy or radio emission was used to find the
most probable far-IR source for optical spectroscopy.
Even with the broadened far-IR color criteria, we still find that
half of our identifications are spectroscopically classified as
AGN, and some of the remainder have ambiguous emission-line
classifications due to substantial reddening.

\section{Sample and observations}

\subsection{Candidate selection}

We constructed initial source lists using the data holdings at
Leiden. Each set of overlapping AO scans (or individual scans
with no overlap) in the 1040 fields centered at galactic latitudes
$\strut b \strut > 20^\circ$ was coadded in the sky plane, extracting sources
using the Westerbork algorithm. In this algorithm, sources are identified
based on number of pixels above a set threshold in signal-to-noise ratio 
(typically $5 \sigma$, with fluxes evaluated by fitting a template
point-spread function centered on these source positions.
These source lists
were filtered for transient glitches by comparison of individual
scans wherever multiple observations were taken; sources where
no such confirmation was possible were flagged (NC in Table 1).
Color selection used the fluxes extracted with a point-source template from
the coadded data. In some cases, we could derive improved flux data
by incorporating {\it scanpi} processing of co-added survey data,
where a large number of passes were available at high ecliptic latitudes. 
Our color criteria were based on the PSC AGN
results (de Grijp et al. 1985, 1987), extended to include typical 
AGN spectra at redshifts
up to $z=1$. Expressed in spectral indices among the IRAS bands,
they included the ranges $\alpha_{25,60} = [-1.5, +0.5]$ or
$\alpha_{60,100} = [-0.8,+0.5]$ where spectral indices are given
in the convention $F_\nu \propto \nu^\alpha$, so that F$_\nu$ dropping to
longer wavelengths means $\alpha > 0$. These spectral indices are
evaluated simply from the in-band fluxes with no finite-passband
corrections, whose effect stretches the color space at extreme values.
These ranges are designed to exclude stellar photospheres ($\alpha \approx 2$)
and as much as possible of the cold dust emission
from ``normal" galaxies. 

To allow for sources near the detection limit,
we now accept sources with an unusually ``warm" spectral shape from
either $12-25 \mu$ or $60-100 \mu$ even when there is no detection
in the adjacent passband.

This effort was parallel to, but independent of, the
IRAS Serendipitous Survey Catalog (SSC, Kleinmann et al. 1986). There
is substantial overlap in the objects included with both processing
streams, but differences in data handling meant that some objects
passed our color criteria in only one survey; for completeness,
we incorporated objects which
passed our color thresholds as measured by either project. 

We adopted the position measured at the shortest wavelength with a
significant detection. However, the IRAS AO positions are inherently
less certain than those from the PSC, for two reasons. The shorter
arcs of these observations mean that the entire orbital attitude
solution from stellar transits could not be used to refine the
spacecraft attitude. Furthermore, attempts to measure offsets
for each field using bright stars detected at $12 \mu$ showed that
sampling issues limit the accuracy of these positions to about the
same level (30--60", depending on scan direction). 
After the fact, using well-established optical identifications,
objects also included in the IRAS Faint-Source Survey (FSS;
Moshir et al. 1992),
and radio measurements, we find that a typical error ellipse has radii of 
about $30 \times 60$
arcseconds, longest in the cross-scan direction (roughly parallel to
lines of ecliptic latitude). This error ellipse contains about 85\% of
spectroscopically identified emission-line galaxies and
radio-IRAS cross-identifications.
 
By convention (Keel et al. 1988), these sources were designated as
IRAS-L, using truncated 1950 coordinates as in the example 
IRAS-L 08200+1055, a Sy 1 reported in our earlier paper. 
Our candidate list is given in
Table 1, with the IRAS fluxes and spectral fluxes. The origin of each
source's data, Leiden or Serendipity Survey processing, is given by the
code L or S; the appropriate source designation may be derived from
this and the listed coordinates, kept at epoch 1950
in this table for consistency with the original
measurements. We designate with S those sources which satisfied our
search criteria in the SSC processing but not the Leiden processing;
upper limits were not listed for the SSC results.
Because of infrared cirrus and detector flashing, useful upper
limits are not uniformly available for nondetections, signified by a zero 
in the flux listing and
omission of the spectral indices involving the affected bands.
A few rich Magellanic Cloud fields and reflection nebulae
are omitted here; the list includes some potential stellar
identifications discussed in Appendix A. In addition to the new
objects in Table 1, we recovered 61 known
AGN in this study, either radio or Seyfert galaxies that were themselves the
targets of the AOs or other objects already found from the Point-Source
Catalog by de Grijp et al. (1987, 1992). We have attempted to remove from
our sample any galaxies which were themselves the target of the AO
pointing, although this took some guesswork in a few cases where
early IRAS documentation did not record the reason for a pointing.

One region in Hercules was especially well-covered by multiple AO
scans, affording a set of far-IR detections to unusually deep levels.
As a reconnaissance of the deep far-IR content of the sky, we have examined
the identifications of all the IRAS sources in this field regardless of
colors, as described in Appendix B.

\subsection{Optical Spectroscopy}

Optical spectra of candidates, sometimes several for a single IRAS
source, were obtained with several instruments. In most cases,
the data could be placed on a flux scale through observations of
standard stars.

We used the 2.5m Isaac Newton Telescope (INT) with Intermediate-Dispersion
Spectrograph (IDS) in two modes, a moderate-resolution mode using the
Image Photon-Counting System (IPCS) as detector with realtime readout,
and the higher-throughput Faint-Object Spectrograph (FOS) with
coarser spectral sampling. With the IPCS, the range 3200-7100 \AA\ 
was sampled with 2040 pixels covering 2.0 \AA\  each. The FOS
data cover 5000-10000 \AA\ , with orders separated by a cross-dispersing prism.
Their resolution is set by sampling at 2 pixels, or 21 \AA\ .

Most of the southern candidates were observed using the ESO Faint-Object
Spectrograph and Camera (EFOSC) at the ESO 3.6m telescope. The acquisition
images from this instrument serve for the photometric and morphological
properties of many of these targets. Most data were taken with the
``blue" grism (B300, 3800-7000 \AA\ ). 
A 1.5" slit gave spectral resolution of 13 \AA\  FWHM
with spectral sampling of 6.9 \AA\ .

With the 3.8m Mayall telescope of Kitt Peak National Observatory, we used
the Cryogenic Camera grism system. A 2.5" slit gave resolution for
a filled slit of 16 \AA\  
FWHM across most of the 4500-8000  \AA\ spectral range covered by
grism 770, somewhat degraded at the ends by detector focus. For most
objects, nuclear spectra have somewhat better resolution, especially
important in deblending the H$\alpha$ and adjacent [N II]
lines. Six final objects were identified in 2004 using the MARS
enhancement of the Cryogenic Camera, and a 1.8" slit, observing the
4500-10300 \AA\  range essentially free of focus changes. This
final subset was not included in the statistics of positional offests,
since they were identified with knowledge of the radio positions.

Observations at the 4.2m William Herschel Telescope (WHT) used the Faint-Object
Spectrograph, with a fixed spectral range of 4600-9500 \AA\  in the
first order. A prism cross-disperser removes potential
second-order contamination. These data suffered from electronic crosstalk,
manifested as a low-level additive Moire pattern, which we reduced by fitting
local sine waves in the Fourier domain to the interference. Telluric
A and B bands were approximately corrected using observations of the same
hot standard stars used for flux calibration. The 1.1" slit
gave a spectral resolution limited by Nyquist sampling (2 pixels
correspond to 17.4 \AA\ ).
          
A few objects were observed using the GoldCam spectrometer at the
Kitt Peak 2.1m telescope. A Loral 3K$\times$ 1K CCD covered the
range $4460--8300$ \AA\  with 1.25-\AA\  sampling, at a resolution
of 3.8 \AA\  FWHM set by the 2.0" slit width.

Some particularly
ambiguous fields were initially observed using slitless spectroscopy
at the ESO 3.6m or KPNO 4m telescopes, so that emission-line
objects would reveal themselves across the entire IRAS position box.

These optical spectra give us secure identifications and redshifts 
for 74 of the 180 new candidates.
Of these, 35 (47\%) are clear active nuclei with Sy 1, Sy 2, or strong LINER
spectra. The remained include definite starbursts, some strongly 
reddened, and objects of ambiguous nature for which we detect H$\alpha$,
[N II], and usually [S II] but none of the bluer lines. Many of these
might be either strongly obscured AGN or star-forming systems. The spectroscopic
results are given in Table 2. In two cases, 09426-1929 and
17435+3141, we find multiple emission-line
objects at nearly the same redshift, and assign the IRAS identification
to the object with the strongest line emission.
In many cases, we could classify the spectrum using ratios of the
strong emission lines, following the precepts of
Baldwin, Phillips, \& Terlevich (1981) and Veilleux \& Osterbrock (1987).
These classifications are given in Table 2. 
For cataloguing purposes, we list the optical coordinates of these
objects, as obtained by centroiding around the peak from Digitized Sky
Survey DSS2 red-light images.
%Line fluxes are given as calibrated 
%using standard stars with the same slit width, and line ratios
%were measured using Gaussian deblending where necessary.

Spectra of the unambiguous AGN are shown in Fig. 1, except for the
composite emission/absorption object 1226+32 (section 4.2). 

\subsection{Optical imaging}

In support of the identification program, as well as providing fluxes
and morphological information, we obtained $R$-band CCD images of most of the
candidates from Table 1. These were done between 1986 and 1992 using
telescopes at Lowell Observatory, ESO, and Kitt Peak. These data are used 
to illustrate the identifications (Fig. 2). In each case, the IRAS
position is shown by an error ellipse $60 \times 120$ arcseconds in size
at the appropriate orientation. The spectroscopic or radio identification
is marked wherever known. In this atlas, images not otherwise
labelled are from the 1.1m Hall telescope of Lowell Observatory.
The new imaging is nearly complete north of $\delta = -25^\circ$,
and has been supplemented for illustration by red-light images 
(where available) from the Digitized Sky Survey. Where the identification
is clear from spectroscopic or radio data, we present $R$-band magnitudes
of these sources in Table 3. These are mainly intended to guide followup
observations; because of the wide range of angular sizes, they may have
errors $\pm 0.2$ magnitude from a consistent definition of either
isophotal or total values. These objects range from $R=12.0-20.5$,
with median 15.7 and quartile points at 14.7 and 17.2. The source counts
for objects with secure optical identifications depart strongly from
a Euclidean slope for $R>15.5$, indicating that we are still missing some
even at these bright levels. This is certainly due in part to
incomplete spectroscopic data.

\subsection{Radio fluxes}

The strong correlation between far-IR and radio fluxes for both active
nuclei and star-forming systems suggest that we can refine the positions of
candidate AGN using radio data at higher spatial resolution. We initially
observed 33 of these fields using the NRAO\footnote{The National Radio 
Astronomy Observatory is a facility of the National Science Foundation operated 
under cooperative agreement by Associated Universities, Inc.} 
VLA at 6 cm. 
After most of the optical observations were finished, the VLA sky surveys 
(FIRST and NVSS) became
available, allowing us to refine the positions and nature of many additional
sources. The radio properties, from our own and the survey observations, are
given in Table 4. The radio position is taken from the new 6-cm data,
the FIRST (Becker et al. 1995) catalog, and the NVSS (Condon et al. 1998), 
in that order of precision.
A correspondence is counted as a counterpart if it lies
with 80" of the IRAS position. Based on the surface density of FIRST
sources at larger projected radii, we expect $<3$ sources out of our 67
radio detections to be chance coincidences
within this radius, and the statistics of position matches suggest that
very few of the IRAS positions are farther off than this.
Locations of the radio-continuum sources are shown by circles in Fig. 2.
These coincide with plausible optical galaxies for an additional 12
high-probability identifications for which we do not 
have optical spectra. There are also another 12 fields where an
optical galaxy is so bright as to be a likely identification, but
for which we do not have spectroscopic data.

Our new VLA observations used the D array at 6 cm. We targeted
26 IRAS fields, with position and flux results included in Table
4. Among these, there were ten nondetections; 1704+606 is not
a sensitive limit, being lost in the sidelobes of 3C 351. Among
the detections, most are unresolved using the D array; a few 
(1439+354, 1712+098, 2028-128) are resolved with Gaussian-fit FWHM in 
the range 9-36".

In two of the 42 cases with both kinds of identification available, the 
emission-line galaxy is not the same object as
the radio source. It is not clear whether the true far-IR sources
have only weak line emission, or whether these are the two chance radio sources
that would be expected in this sample. These are 
1512+366, where the radio source has $R=18.84$, while there is an
brighter emission-line galaxy at the edge of the IRAS error ellipse; and
1503+272, with comparably faint spectroscopic and radio identifications
at opposite sides of the IRAS ellipse.

\section{Sample properties}

Comparison of the redshift and magnitude distributions with earlier
samples can clarify the ways in which these objects extend our
knowledge of the far-IR Universe. In particular, we expect this
sample to correspond roughly to the color-selected AGN found
from the Point-Source Catalog, shifted to larger redshift as
a result of the deeper flux limit.

The redshift distribution of the spectroscopic identifications ranges
up to $z=0.51$, with quartiles at $z=0.035, 0.112$ and median $z=0.067$.
For the Seyfert galaxies, the quartiles are $z=0.045,0.11$, with the
median and high values the same. Of the entire sample, 10\% lie at
$z>0.23$, while 10\% of Seyfert galaxies are at $z>0.18$. For the
Seyfert galaxies in the Point-Source Catalog sample (de Grijp et al. 1992),
the full range is $z=0.0026-0.91$, with median $z=0.0383$ and quartiles
$z=0.022,0.061$. Of these, 10\% lie beyond $z=0.11$. Three of the
226 PSC AGN lie beyond our highest redshift at $z=0.53$, which is
consistent with the much larger PSC sample size and our
exclusion of bright, previously known objects from
the AO sample. For all 370 extragalactic
objects in the PSC ``warm" sample, the redshift range spans $z=0.0009-0.91$,
with median $z=0.035$ and quartiles $z=0.02,0.056$; 10\% of the whole
list has $z>0.092$. The cumulative distributions in redshift are best matched 
if the AO sample of Seyferts reaches on average $1.8 \times$ deeper than
the PSC Seyfert, which implies a typical difference of $3.2 \times$ in
flux. This confirms that, to first order, the new sample is reaching
farther rather than to lower luminosity. The uneven sampling depth
means that we see about $4 \times$ the number of warm sources
that simple scaling from the PSC predicts for the mean depth and
area covered.

At each redshift, our criteria include only
a fraction of Seyfert galaxies, since they have an intrinsic spread
in IR colors and a varying mix of host-galaxy dust emission as well.  
We can evaluate this fraction at small $z$ directly, using the samples 
of Markarian and CfA Seyferts (Lipovetskii et al. 1988, Huchra \& Burg 1992), 
with fluxes taken from the IRAS Faint-Source Catalog (FSC). 
Of the 205 Markarian Seyfert galaxies,
149 were detected by IRAS, of which 110 fall within our color bounds.
Nearly half (57) are listed as Sy 1, with the remainder almost evenly
split between ``intermediate" types 1.2--1.9 and Sey 2. Among the
detected Markarian Seyferts, there is strong overlap in
FIR luminosity between those included and excluded by our color
criteria. The CfA Seyfert nuclei are generally intrinsically fainter,
and only 31 of 86 pass these color criteria. Of the remainder, 15 were 
not detected at all by IRAS. Among these, there is a strong color
change with FIR luminosity; the median L(FIR) for objects that would
be included in our sample is four times higher than for the ones we reject,
which fits with an increasing role for the surrounding cold dust from the
host galaxy in fainter nuclei. Sy 1 nuclei slightly outnumber Sy 2's
among the ``warm" objects in the CfA sample (15:10), comparable to
the ratio (13:17) seen in our newly-found AGN. This surely
reflects a bias in the contrast of the AGN emission against the
surrounding host galaxy; Alexander (2001) has found that the far-IR
spectral shape mostly reflects the ratio of nuclear and disk
emission. Some additional Sy 2 objects may be hiding within the
galaxies for which we see only the red emission lines, limiting
our ability to classify them.

Of the 94 optical and radio identifications, 13 have probable
companion galaxies, while another 13 show disturbed morphologies
in our images. These are both quite high fractions compared to
optically-selected samples, but are not exceptional for
samples selected at high far-IR luminosity (see, for example,
Gallimore \& Keel 1993 and references therein).

\section{Active Galactic Nuclei}

\subsection{Sample properties}

As illustrated by the spectra in Fig. 2, the active nuclei in this
sample show a range of spectroscopic properties, from classic
Fe-rich Sy 1 to strongly obscured objects with only the red
emission lines from a narrow-line region detected. Previous
infrared-selected samples show large proportions of Sy 2 objects,
since the strong blue continuum of typical Sy 1 no longer introduces a
direct selection bias. For example, the sample listed by de Grijp et al. (1992) 
from the Point-Source Catalog includes 144 Sy 2 and 82 Sy 1 objects.
The deeper AO sample fits this trend within
our modest statistics. Including
Sy 1 with the broad component detected only at H$\alpha$, we find roughly
equal numbers of Sy 1 (13) and Sy 2 (17) among our identifications.

Several previous studies have addressed the demographics of AGN selected
from colors in the IRAS Point-Source Catalog. The objects listed by de Grijp et 
al. (1992),
selected for $\alpha (25,60)$ in the range [-1.5,0], show a preponderence
of Sy 2, finding 141 Sy 2 and 80 Sy 1. Our current sample shows
a Sy 2:Sy 1 ratio of 17:13. Using the same
25-60$\mu$m criterion on a limited sample from the Faint-Source Catalog,
Gu et al. (1995, 1997) find nine Sy 2 and no Sy 1. Low et al. (1988),
using $/alpha (25,60) < 1.25$, found a high ratio 73:39. These values
are broadly consistent with a fraction of Sy2/(Sy1 + Sy2) near 0.65,
within binomial statistics from these samples. The sole possible outlier
is the small Gu et al. sample, which lies about $3 \sigma$ from this
common mean value (having no broad-line objects). It is thus a general
feature of far-IR color selection that roughly twice as many
type 2 as type 1 AGN appear, reinforcing previous studies in the
number of narrow-line AGN which are missed by many optical surveys.
This stands in contrast to the dominance of broad-line objects in
samples selected for red $J-K$ colors, consistent with the small fraction of
our objects falling in this category.

Some studies have indicated a connection between the far-IR spectral
shape of AGN and the strength of the optical Fe II emission multiplets
(Lawrence et al. 1988; Low et al. 1989; Lipari, Macchetto, \& Golombek 1991;
Zheng \& Keel 1991).
This sample includes a few individual AGN with very strong Fe II.
Four of these are noted below.

A few of the obvious Seyfert nuclei in this sample are highly reddened,
as shown by broad H$\alpha$ and little or no H$\beta$ emission. There
are probably additional reddened Sy 2 galaxies hidden among the
ambiguous objects for which we do not detect enough emission lines
to make a definite classification, since the range of [N II]/H$\alpha$
for Seyfert 2 and starburst nuclei overlaps.

Of the objects with clear radio or optical counterparts, 33 are detected in 
the 2MASS survey catalogs. Taking the extended-source entries for
objects which also appear in the point-source listing, and combining
the near-IR colors with their $R$ magnitudes, shows clear relations
between $J-H$ or $H-K$ and $R-K$ (Fig. 3). Objects without spectroscopic
identifications (either due to faintness or lack of emission lines)
are redder in $R-K$ than the entire sample, while spanning similar
ranges of JHK colors. This is at least consistent with some of the
IRAS AO sources without an optical identification 
having strongly obscured nuclei, either starbursts or AGN. Among
objects with spectroscopic identifications, starbursts are bluer
across the JHK bands than the few Seyferts detected; some Sy 1 are
as red as $J-H=1.1, H-K=1.3$. However, this far-IR color selection
does not (yet?) uncover many of the kind of red AGN identified by Cutri
et al. (2000); among the four 2MASS detections of Sy 1 in our sample, only 
22212-0221 satisfies their $J-K > 2$ criterion. To be consistent
with their measurement of the space densities of type 1 AGN redder and
bluer than this limit would require that either the reddest AGN in our sample
remain preferentially unidentified, or that red AGN generally exhibit
cooler far-IR spectral shapes.

\subsection{Individual noteworthy objects}

IRAS-L 06229-6434 - The continuum is flat in F$_\lambda$, with very strong Fe II 
emission.

IRAS-L 08020+1055 - The Sy 1 classification is based on the width of H$\alpha$,
since H$\beta$ and [O III] are too weak for detection.
 
IRAS-L 11549+7327 - This is the highest-redshift object we have identified,
at $z=0.51$. Its continuum is very blue, and Fe II emission is
strong on either side of H$\beta$.

IRAS-L 12266+3240 - This composite object includes narrow emission lines at a
wide range of ionization, as in a LINER or Sy 2, superimposed on the
starlight of a fading starburst. This is seen most clearly through the
strong absorption at H$\beta$ and H$\gamma$(Fig. 4). 
There is a companion projected
at 3.5" (10 kpc at $z=0.17$). The stellar population appears similar
to that in the nucleus of NGC 4569 (e.g., Keel 1996), in which
A-type stars are dominant in the optical.

IRAS-S 14400+3539 - Fe II is strong.

IRAS-L 19051-6233 - The continuum is blue, and Fe II is very strong.

IRAS-S 22212-0221 - The broad Balmer lines are very strong, while Fe II emission 
is rather weak. This object has the reddest JHK colors among AGN in this
sample also detected in the 2MASS survey.

\section{Star-forming galaxies}

Selecting galaxies for unusually warm far-IR colors isolates some
kinds of star-forming objects as well as active nuclei. Many of these
are optically very blue. The apparent paradox of dusty galaxies
not always appearing reddened has been discussed by Witt, Thronson, \&
Capuano (1992) and by Meurer et al. (1995). These authors (among others) point out
that when dust and stars are mixed, reddened stars are effectively removed from
the observed mix at short wavelengths, so the emerging radiation can be blue (although
attenuated from the dust-free case). Scattered light will only accentuate this
effect. 

A few of these sources are worthy of individual note. In IRAS-L 
09426-1929, both members of this strongly interacting pair, MCG -3-25-19 or Arp 252,
likely contribute to FIR. The northern component dominates at 20 cm.
Both
17235+4602 and 10360-0643 (IC 630, Mkn 1259) have blue continua and strong
Wolf-Rayet N IV + He II + C II blends near 4600 \AA\ .

\section{Blank fields - faint objects or statistical mirage?}

Some of the IRAS sources in our sample have no plausible optical
identifications to $R \approx 20$. If real, these are particularly
interesting as candidates for high-redshift or heavily-obscured objects.
There is precedent for far-IR sources so luminous that we
could detect them to high redshifts if we were lucky enough to
have one in these fields - the central cluster galaxy with obscured
AGN in 09104+4109 appears in the Point-Source
Catalog (Kleinmann et al. 1988) at $z=0.44$, and the lensed
object from the Faint Source Survey, 10214+4724, is at
$z=2.3$ (Rowan-Robinson et al. 1991).
However, as we go deeper into the IRAS data, there are also concerns
about the reality of these sources, especially those seen in only
a single scan. The best examples of potential blank-field
sources are 02247+1017, 04114+1150,
and 13557+6950, each of which passed the multiple-detection test
on overlapping IRAS scans.
A few other sources have only very faint potential optical counterparts
($R \approx 20$), such as 17467+5600 and 22511-1805; this last one
was detected on only a single scan. These objects are not detected in the
VLA surveys, except for a possible counterpart of 22551-1805 in the
somewhat confused field of MR2251-18.

There are also a few sources in which we found only absorption-line
galaxies - both members of the galaxy pair in the IRAS error
ellipse for 09156+4213, and the brightest galaxy near 09189+4620.
If these are the correct identification for the infrared sources,
the cores of these galaxies may be very heavily obscured; and
if not, the genuine IR source must be substantially fainter.

Sources with no multiple-observations confirmation (labelled with
NC in Table 1) do not seem to be any less reliable than the
confirmed sources. Their statistics of spectroscopic and radio
identification, and of both together, are in fact better than
for the sample as a whole. Of 19 NC sources, we have spectroscopic
emission-line identifications for 10 and radio identifications for
11 (of which nine are in common). Omitting two Galactic objects
and an additional bright galaxy without an observed spectrum, there are
only 4 unidentified NC sources (21\%), compared with 43/180 (24\%)
of the overall sample for which no optical identification is
apparent (including bright galaxies not yet observed in detail).
The remaining 4 NC sources are thus good candidates for genuine
optically-faint warm IR sources.

\section{Summary}

We have selected warm far-IR sources from the deep IRAS Additional
Observation (AO) data, using color criteria designed to preferentially select
warm AGN at redshifts $Z < 1$ from the source population. Of 180
such sources taken from 1040 fields covering $\approx 780$ square degrees,
we have obtained emission-line identifications for 74 (41\%),
while VLA survey observations suggest identifications for an
additional 20 objects without extant optical spectroscopy. Compared
to analogous samples from the IRAS Point-Source Catalog, this
sample goes farther in redshift (rather than to lower luminosities)
as expected for the deeper flux limit. The median redshift for
the entire sample and for identified Seyfert galaxies is identical
at $z=0.066$. 

The population of AGN selected in this way includes numerous objects
with strongly reddened emission-lines, which may be increased by
a class of galaxies in which we see only narrow H$\alpha$+[N II]
emission, which could be either reddened AGN or dusty starburst systems.
We also find several Sy 1 nuclei with very strong Fe II emission,
and one composite object with AGN-like emission lines superimposed
on a bright, fading starburst population. Likewise, the star-forming
galaxies in this group span the full range of optical color from
virtually unreddened to highly obscured. Similar objects in the Point-Source 
Catalog suggest that the youngest starbursts systematically have
the warmest far-IR dust distributions (Fehmers et al. 1994), which
may fit with the fact that two of the brightest of these objects,
among the few with spectra good enough to tell,
exhibit strong Wolf-Rayet blends near 4650 \AA\ .

\section{Appendix 1: Stellar Source Identifications}

For 10 of these warm IRAS sources, the most obvious optical identification
is with a bright star, and for one fainter one, 13446-3625, the
sole emission-line object we have identified has the composite spectrum
of a symbiotic star. We summarize here the properties of these
stellar IR sources.

A sufficiently bright star has a high probability of being the
correct IRAS source identification simply on grounds of low surface
density. Based on latitude-averaged star counts, we expect only a single 
star brighter
than $V=11.0$ by chance within all 180 of the $30 \times 60$"
error ellipses in our survey area. For brighter
stars, identifications are quite secure; for example, the chance of
a single random star with $V<6.0$ occuring within any of our
search ellipses is $< 0.01$.

There are 14 stars with $V < 10.5$ in our sample. Three are catalogued
semiregular variables of late spectral type for which circumstellar
envelopes would be expected (and these may have been the original
targets of the pointed observations). These are 03536-7411 = UY Hyi,
16308-1601 = T Oph, and 09371+3620 = Z LMi. Similarly, 05142-0154 is
the T Tauri star HD 290172.
One additional fainter
star, 13446-3625, turned up in slitless spectra and shows the
characteristic combination of a cool stellar continuum plus
both low- and high-ionization emission lines for identification
as a symbiotic star (Fig. 5). This combination of properties allowed
it to masquerade as a Sy 2 galaxy until slit spectroscopy showed
just how sharp the TiO bandheads are, and a radial velocity
$\approx -40$ km s$^{-1}$.

A few of these stars have been examined from the IRAS all-sky data,
although their AO data do not seem to have been hitherto published.
Reddy \& Parthasarathy (1996) used photometry and optical spectroscopy to 
classify
05239-0626 (BD $-6^\circ$ 1178) as an F2 II star, in a post-AGN phase
in which recently ejected dust gives the FIR emission. Patten \& Willson
(1991) include 05356-2839 (HD 37484) as an (otherwise normal)
IR-excess star based on its
12$\mu$ flux from the Point-Source Catalog.

The remaining stellar identifications include possible examples of the
``Vega phenomenon", reradiation from large grains perhaps representing
the remnant of a protoplanetary disk. They are listed in Table 5, with
identification information from SIMBAD and distances from
{\it Hipparcos}. Among these, 54 Oph is a double with
the brighter component tabulated. These stars are unusual in extending to
later spectral type than usually associated with protoplanetary disks.

\section{Appendix 2: Source Identifications in a Deep Survey Field}

The IRAS AO data include a very deep set of scans of a region in
Hercules near (1950) 1632+3815, outlining a roughly N-S field 1.5$^\circ$
in length. This region, about 2$^\circ$ NW of M13, has an unusually
low level of IR cirrus emission for its galactic latitude,
allowing a very deep set of source extractions and identifications,
reaching to 0.01 Jy at 60$\mu$.
We have considered the nature of all 31 sources identified in this
field (Table 6) as a probe of faint FIR populations in general. Of these,
only two (both SAO stars) are found in the Faint-Source Catalog. The original
CRI data products are no longer available, so we have assessed the
noise level using the gridded IRAS products from the Astronomical Data
Center, adding the 8 constituent observations. This gives $1-\sigma$
point-source errors 0.004-0.006 Jy, so that only the weakest
source we consider is as faint as 3$\sigma$. There
is enough cirrus emission for the detection threshold to change
across the field. We
rely mainly on the 60$\mu$ source list; only three additional sources
might be real at 100$\mu$. At 25$\mu$, the only detections are
the two Seyfert galaxies, SAO stars 1 and 6, an additional star 3,
and source 11. The numbering is consecutive N-S among 60$\mu$
detections. 

We obtained $R$ images of all these sources except the brighter SAO
star, equally divided between the KPNO 2.1m (TI CCD, binned at
0.39"/pixel) and Lowell 1.1m (TI CCD with focal reducer, 0.71"/pixel).
Eight sources with obvious optical counterparts were examined spectroscopically 
with the WHT and Faint-Object Spectrograph (section 2.2). In-scan position 
uncertainties yield error ellipses alongated E-W in this field. The
sources are listed in Table 6.
Two of them are Seyfert galaxies, both of which are bright enough 
and spectrally warm enough to be included
in our main all-sky sample. Source 4 in this field is
IRAS-S 16320+3730, while source 8
is IRAS-L 16327+3747. In addition, source number 2 is W Herculis, 
an OH/IR Mira-type star whose dst properties have been
derived from the IRAS PSC by Hashimoto (1994).

As is the case for the main AO AGN sample, we can improve the quality of source
identifications by exploiting the strong connection between far-IR
and radio flux. Eight of these IRAS sources have a FIRST 20-cm source
within 90"; in one case, there are two FIRST sources, one associated with
a bright disk galaxy and the other a blank field. IRAS source 20 is
associated with the blazar 1633+382 (B2 1633+28, 4C +38.41) at
$z=1.814$, showing strong radio emission with a 20-cm flux density
of 2 Jy. The source identifications include a starburst galaxy at 
$z = 0.0974$ (number 9)
and a bright outlying member of a rich group or cluster (number 15). Our
initial identification in that case was the central cD galaxy, which
suggested an analogy with IRAS 09104+4109 (Kleinmann et al. 1988), but 
the radio source clearly belongs to the secondary elliptical.
These source associations are of high significance; the
source density above 2 mJy in this region of the FIRST survey
would give 1 unrelated source within the sum of 31 IRAS error ellipses.
Some of the optical identifications with FIRST coordinates are quite
faint, as might be expected from the low IRAS flux levels we
reach in this field. As a result, to the extent that the radio-infrared
identifications are identical at this flux level, we were less
successful in guessing the proper objects for spectroscopy based
on the IRAS positions and optical images alone than for the stronger AGN sample.
This underscores the importance of accurate coordinates (often assisted
by the radio-infrared correlation) in identification of such faint infrared
sources. 

\acknowledgments
The data handling and source extraction used the facilities of the Centraal
Rekeninstituut Leiden. The early stages of this project benefited from the
encouragement and support of H. van der Laan. IRAS was a joint project of NASA, 
the UK Science and Enginerring Research Council, and the Netherlands
Agency for Aerospace Programs (NIVR). We thank the successive
directors of Lowell Observatory, Jay Gallagher and Bob Millis, for
important time allocations. Mrs. Heidi Versteeg carried out the checks for
duplicate detections from multiple observations. This work has
benefitted from NASA's {\it SkyView} facility
(http://skyview.gsfc.nasa.gov) located at NASA Goddard
Space Flight Center, and from the rich harvests of data provided by the
VLA FIRST and NVSS surveys.
We thank Peter Allan for providing software to help us unscramble WHT Starlink 
data far from the UK, and Roque de los Muchachos mountain staff for
their work in recovering operations after the summit was threatened
by wildfire, in time for us to use the entire WHT observing run.
This research has made use of the NASA/IPAC Extragalactic Database (NED),
which is operated by the Jet Propulsion Laboratory, Caltech, under
contract with the National Aeronautics and Space Administration.
This publication makes use of data products from the Two Micron All
Sky Survey, which is a joint project of the University of Massachusetts and
the Infrared Processing and Analysis Center/California Institute of
Technology, funded by the National Aeronautics and Space
Administration and the National Science Foundation.
The Digitized Sky Survey is based on photographic data obtained using 
the Oschin Schmidt Telescope
on Palomar Mountain.  The Palomar Observatory Sky Survey was funded
by the National Geographic Society.  The Oschin Schmidt Telescope is
operated by the California Institute of Technology and Palomar
Observatory.  The plates were processed into the present compressed
digital format with their permission.  The Digitized Sky Survey was
produced at the Space Telescope Science Institute (ST ScI) under
U. S. Goverment grant NAG W-2166.
Thanks to Lisa Frattare for help with some of the imaging observations
at Lowell.

\clearpage
%Fig. 1:
\figcaption
{Identification charts for warm sources from the IRAS AO data. Each error
ellipse is $60 \times 120$ arcseconds, oriented with the short axis
in the better-determined in-scan direction. Unless otherwise labelled,
images are $R$-band data from the Lowell 1.1-m Hall telescope; only two
fields (showing DSS O data) are not from red-light images.
Verrtical and horizontal tick marks indicate spectroscopic identifications,
and circles indicate positions of radio sources. Each field shown is
$3 \times 3$ arcminutes with north at the top. In a few fields, IPCS
spectra showed no emission lines in bright galaxies; these are indicated. 
\label{fig1}}

%Fig. 2:
\figcaption
{Spectra of identified AGN, all shown in the observed wavelength frame,
and marked with the emission-line classification. One object, IRAS-L
17544+4347, a Seyfert 2 at $z=0.07288$, was observed after figure preparation 
and is not plotted, while the composite spectrum of IRAS-L 12266+3240
is shown in detail in Fig. 3. The flux scale is in units of 
$10^{-16}$ erg cm$^{-2}$ s${-1}$ \AA\ $^{-1}$.
\label{fig2}}

%Fig. 3:
\figcaption
{JHK colors, from the 2MASS catalog releases, as a function of $R-K$
between our CCD measures and the 2MASS data. Symbol types distinguish
classifications based on the optical spectrum. We detect only a single
object which satisfies the definition of a ``2MASS red AGN". Note that
the yet-unidentified objects, with no firm spectroscopic identification,
are systematically redder in $JHK$ than the identified objects, which
could mean that some of them represent heavily obscured active nuclei.
While there is a general relationn between $R-K$ and the $JHK$ colors,
enough outliers appear to suggest that host-galaxy starlight dominates
the flux in many objects. 
\label{fig3}}
newfig 3 JHK

%Fig. 4:
\figcaption
{The composite AGN+post-starburst spectrum of IRAS-L 12266+3240
at $z=0.173$. The WHT FOS and higher-resolution KPNO Cryogenic Camera
(boxcar-smoothed to resolution of 12 \AA\ ) data are superimposed. Features
of a stellar population several $10^8$ years old dominate, with strong Balmer
absorption at H$\beta$ and shortward. Both datasets appear to show
a blue component in the Na D lines, a potential signature of a
strong high-velocity starburst wind, but this spectral region would br
contaminated by incompletely corrected
telluric B-band absorption, so the feature
may not be astronomical in origin. For comparison, we show the spectrum of
NGC 4569 (shifted downward by 0.7 units), which is well fitted
by a $10^8$-year burst for the optical spectrum (Keel 1996). Important
emission and absorption features are marked, with telluric absorption
and night-sky residuals indicated by the ``earth" symbol.
\label{fig4}}

%Fig. 5:
\figcaption
{ESO 3.6 EFOSC spectrum of the symbiotic star IRAS-L 13446-3625.
The symbitic classification is driven by the large ionization
range in the emission lines, with [O I], [O II], and [O III]
comparably strong. The cool stellar contribution is especially
marked in the sharp TiO bandheads. Important emission and absorption
features are marked, as well as night-sky residuals at 5577 \AA\ .
\label{fig5}}

\clearpage

\makeatletter
\def\jnl@aj{AJ}
\ifx\revtex@jnl\jnl@aj\let\tablebreak=\nl\fi
\makeatother
\begin{deluxetable}{cccccccc}
\tablecaption{IRAS Properties of Program Sources}
\tablehead{
\colhead{$\alpha_{50}$}           & \colhead{$\delta_{50}$}      &
\colhead{List}          & \colhead{F$_{12}$}  &
\colhead{F$_{25}$}          & \colhead{F$_{60}$}    &
\colhead{F$_{100}$}  & \colhead{Conf?} }
\startdata
 00 04 12.0 & -55 21 41  & L  & 0.05    & 0.08 &   0.09  & 1.28 \nl                
 00 05 06.2 &  06 16 26  & L  & 0.13    & 0.29 &   0.25  & $<0.25$ & NC \nl                
 00 16 48.4 &  26 37 42  & L  & $<0.04$ & 0.16 &   0.36  & 1.10    \nl   
 00 17 54.6 &  15 03 34  & L  & 0.00    & 0.11 &   0.36  & 0.00       \nl
 00 21 56.2 & -72 20 54  & S  & $<0.08$ & 0.48 &   0.19  & $<0.45$ \nl
 00 31 53.0 & -41 38 56  & L  & $<0.04$ & 0.10 &   0.75  & 0.92   \nl
 00 32 50.0 & -79 20 07  & S  & $<0.07$ & 0.18 &   0.20  & $<0.41$ \nl
 00 55 38.2 &  29 37 22  & L  & 0.00    & 0.00 &   0.29  & 0.33   \nl      
 01 16 33.3 &  31 46 53  & L  & 0.00    & 0.15 &   0.34  & 0.00    \nl
 01 19 34.4 &  86 49 46  & L  & $<0.03$ & 0.08 &   0.43  & 0.57   \nl
 01 20 08.4 &  25 07 31  & L  & 4.98    & 2.21 &   0.45  & 0.43      \nl          
 01 21 35.3 & -35 19 51  & S  & 0.25    & 0.46 &   0.31  & $<0.60$ \nl 
 01 21 50.9 & -59 04 01  & S  & 0.39    & 0.59 &   0.60  &  0.74         \nl        
 01 41 50.6 & -43 51 54  & L  & $<0.03$ & 0.16 &   0.30  & 0.51    \nl
 01 50 08.4 & -23 34 44  & L  & 0.00    & 0.18 &   0.31  & 0.00  & NC    \nl
 01 58 50.1 & -58 25 37  & L  & 0.10    & 0.20 &   0.21  & 0.48   \nl
 02 03 22.9 &  32 05 07  & L  & 0.00    & 0.00 &   0.29  & 0.35      \nl   
 02 05 15.3 &  02 28 08  & S  & $<0.07$ & 0.13 &   0.10  & $<0.24$ \nl         
 02 20 26.2 &  35 12 38  & S  & $<0.06$ & 0.29 &   0.15  &  0.59         \nl         
 02 24 45.0 &  10 17 48  & S  & $<0.11$ & 2.33 &   1.18  & $<0.57$ \nl   
 02 31 44.7 & -39 05 15  & L  & 0.00    & 0.00 &   0.45  & 0.60   \nl
 02 42 51.5 & -72 30 25  & S  & 0.11    & 0.12 &   0.31  & $<0.45$ \nl
 03 06 13.7 &  10 51 46  & L  & 0.13    & 0.32 &   0.34  & $<0.25$ \nl   
 03 17 59.6 & -72 02 08  & S  & $<0.09$ & 0.13 &   0.40  &  0.65 \nl
 03 19 33.0 & -36 44 48  & L  & 0.00    & 0.04 &   0.14  & 0.00   \nl
 03 53 22.5 &  26 06 19  & S  & $<0.07$ & 0.17 &   0.46  &  1.32 \nl
 03 53 41.0 & -74 11 32  & L  & 11.63   & 4.15 &   1.02  & 0.88    \nl 
 04 11 27.6 &  11 50 38  & L  & 0.15    & 0.28 &   0.21  & $<0.25$ \nl                
 04 14 50.7 & -62 32 10  & L  & 0.00    & 0.04 &   0.14  & 0.80   \nl
 04 21 29.0 & -56 18 44  & S  & $<0.07$ & 0.18 &   0.13  & 1.02 \nl
 04 55 28.8 & -70 31 45  & L  & $<0.04$ & 0.27 &   0.78  & 1.65   \nl 
 04 59 41.4 & -70 35 48  & L  & $<0.04$ & 0.44 &   0.78  & 2.86    \nl
 05 06 18.9 & -61 22 09  & L  & 0.00    & 0.05 &   0.27  & 0.35   \nl
 05 11 36.7 & -33 23 28  & S  & $<0.07$ & 0.11 &   0.23  &  0.36     \nl
 05 14 12.3 & -01 54 42  & S  & 0.98    & 1.32 &   1.00  &  1.22 \nl
 05 18 59.2 & -25 24 36  & L  & 0.70    & 3.44 &  15.64  & 15.30 & NC   \nl 
 05 23 56.2 & -06 26 30  & S  & 1.47    & 1.96 &   1.27  & $<0.61$ \nl    
 05 25 29.4 & -40 10 05  & L  & 0.00    & 0.00 &   0.30  & 0.38  & NC    \nl
 05 25 48.0 & -63 39 20  & S  & 0.16    & 0.36 &   0.18  & $<0.50$ \nl
 05 26 07.2 & -20 40 04  & L  & 0.00    & 0.08 &   0.16  & 0.00        \nl 
 05 35 40.8 & -28 39 10  & L  & 0.136   & 0.08 &   0.13  & $<0.21$ \nl   
 05 56 20.7 & -38 20 11  & S  & 0.52    & 0.70 &   0.39  & $<0.37$ \nl   
 05 57 28.5 & -67 13 38  & S  & $<0.08$ & 0.08 &   0.16  & $<0.39$ \nl
 06 10 12.5 &  70 46 24  & L  & 0.00    & 0.00 &   0.23  & 0.31  \nl
 06 22 55.3 & -64 34 43  & L  & 0.04    & 0.08 &   0.14  & $<0.24$ \nl         
 06 24 18.6 & -53 16 14  & L  & 0.07    & 0.05 &   0.06  & 0.36     \nl              
 06 36 08.3 & -62 17 58  & L  & $<0.03$ & 0.14 &   1.95  & 2.44        \nl  
 06 38 34.3 & -51 22 36  & L  & $<0.09$ & 0.13 &   0.44  & 0.88    \nl
 06 46 49.1 & -74 24 58  & S  & $<0.06$ & 0.12 &   0.37  & 0.55 \nl
 06 55 35.5 &  54 15 57  & S  & 0.10    & 0.20 &   0.32  & 0.62    \nl 
 07 22 39.6 &  31 48 04  & L  & 0.00    & 0.00 &   0.74  & 0.73  & NC     \nl           
 08 02 00.3 &  10 55 00  & L  & 0.00    & 0.15 &   0.50  & 1.48  & NC    \nl
 08 04 27.1 & -71 26 59  & L  & 0.00    & 0.06 &   0.14  & 0.00   \nl
 08 22 48.8 &  29 27 12  & L  & 0.00    & 0.00 &   1.20  & 1.60   \nl
 08 23 17.6 & -77 04 32  & L  & 0.00    & 0.05 &   0.16  & 0.00   \nl
 08 29 15.6 &  04 39 36  & S  & $<0.12$ & 0.33 &   0.42  & $<0.54$ \nl    
 08 33 56.8 &  65 17 43  & L  & 0.26    & 1.33 &   6.73  & 8.63    \nl
 08 37 01.4 &  29 59 17  & S  & $<0.08$ & 0.19 &   0.45  & 0.62     \nl
 08 38 26.5 &  77 04 05  & S  & $<0.08$ & 0.13 &   0.20  & 0.38   \nl
 08 46 53.5 &  19 41 40  & S  & 0.15    & 0.39 &   0.24  & $<0.30$ \nl                  
 08 51 55.9 &  20 18 03  & S  & 0.22    & 0.47 &   0.89  & 1.34    \nl
 09 15 40.1 &  42 13 13  & L  & 0.00    & 0.11 &   0.26  & 0.00      \nl            
 09 18 57.4 &  46 20 02  & L  & 0.00    & 0.03 &   0.11  & 0.00         \nl        
 09 30 21.1 &  22 23 25  & L  & 0.00    & 0.00 &   0.29  & 0.36            \nl      
 09 32 51.4 & -14 36 50  & L  & 0.08    & 0.18 &   0.21  & $<0.15$ \nl   
 09 37 10.8 &  36 20 02  & L  & 2.70    & 1.11 &   0.26  & 0.34               \nl  
 09 42 18.0 & -08 19 24  & L  & 0.00    & 0.00 &   0.28  & 0.36          \nl
 09 42 40.2 & -19 29 09  & L  & 0.15    & 0.87 &   4.24  & 4.81    \nl
 09 44 17.1 & -07 44 53  & L  & 0.00    & 0.00 &   0.34  & 0.34      \nl
 10 11 59.6 & -00 52 36  & L  & 0.29    & 0.11 &   0.10  & 0.33         \nl        
 10 14 57.4 &  21 02 36  & L  & 0.00    & 0.00 &   0.23  & 0.34         \nl
 10 36 02.9 & -06 54 46  & S  & 0.72    & 4.78 &  15.21  & 17.53      \nl
 10 57 06.5 &  72 39 00  & L  & 0.00    & 0.05 &   0.12  & 0.27         \nl
 11 50 46.8 &  58 23 49  & L  & 0.00    & 0.11 &   0.31  & 0.78      \nl
 11 54 54.5 &  73 27 30  & L  & 0.00    & 0.08 &   0.14  & 0.00    \nl
 12 06 15.9 &  68 11 20  & L  & 0.00    & 0.00 &   0.24  & 0.31     \nl
 12 10 50.9 &  37 03 46  & L  & $<0.04$ & 0.16 &   0.59  & 1.09    \nl
 12 17 57.5 &  30 13 31  & L  & 0.00    & 0.00 &   0.19  & 0.27       \nl  
 12 21 30.8 &  11 07 36  & L  & 0.00    & 0.00 &   0.51  & 0.69  & NC    \nl
 12 23 53.9 &  48 46 00  & L  & 0.00    & 0.06 &   0.52  & 0.62    \nl
 12 26 37.8 &  32 40 45  & L  & 0.00    & 0.00 &   0.71  & 1.03     \nl
 12 29 32.4 &  14 13 57  & L  & $<0.08$ & 0.25 &   0.63  & 0.86  & NC    \nl
 12 39 45.5 &  33 33 33  & L  & 0.09    & 0.26 &   0.49  & $<0.21$ \nl    
 12 41 02.3 &  39 23 42  & L  & 0.00    & 0.00 &   0.20  & 0.28      \nl
 12 49 52.2 &  27 21 28  & L  & 0.00    & 0.00 &   0.31  & 0.37      \nl
 12 58 19.4 &  28 47 09  & L  & 0.00    & 0.00 &   0.23  & 0.29     \nl
 13 20 44.8 &  55 10 48  & L  & 0.00    & 0.06 &   0.16  & 0.00  & NC    \nl
 13 44 38.7 & -36 25 11  & L  & 0.34    & 0.78 &   0.78  & $<0.30$ \nl 
 13 55 39.3 &  69 51 08  & L  & $<0.04$ & 0.07 &   0.20  & 0.51    \nl
 14 03 00.7 &  53 38 10  & L  & 0.00    & 0.09 &   0.19  & 0.30  & NC   \nl
 14 07 14.2 &  54 56 51  & L  & 0.00    & 0.00 &   0.23  & 0.33  & NC   \nl
 14 13 34.5 &  13 34 20  & L  & 0.00    & 0.07 &   0.26  & 0.26  \nl
 14 39 01.3 &  35 24 20  & L  & 0.00    & 0.04 &   0.14  & 0.47    \nl
 14 39 59.6 &  36 05 34  & L  & 0.00    & 0.00 &   0.32  & 0.33      \nl
 14 40 04.9 &  35 39 02  & S  & 0.10    & 0.20 &   0.65  & 1.03      \nl
 15 03 21.3 &  27 15 40  & L  & 0.00    & 0.07 &   0.85  & 1.02     \nl
 15 11 16.2 &  11 08 03  & L  & 0.00    & 0.00 &   0.32  & 0.44     \nl
 15 12 36.8 &  36 40 51  & L  & 0.00    & 0.00 &   0.25  & 0.35     \nl
 15 21 39.3 &  30 15 39  & L  & $<0.03$ & 0.05 &   0.11  & 0.34     \nl
 15 26 07.1 &  55 02 17  & L  & 0.00    & 0.00 &   0.60  & 0.78     \nl
 15 34 45.3 &  58 04 10  & S  & 0.11    & 0.24 &   0.21  & $<0.41$ \nl     
 15 35 47.9 &  59 34 08  & L  & $<0.03$ & 0.09 &   0.28  & 0.74     \nl
 15 56 12.6 &  26 00 02  & L  & 0.07    & 0.11 &   0.32  & $<0.19$ \nl    
 15 56 58.6 &  25 40 41  & L  & $<0.03$ & 0.06 &   0.80  & 0.72   \nl
 16 07 32.9 &  29 58 49  & L  & 0.48    & 0.13 &   0.10  & $<0.09$ \nl    
 16 11 52.7 &  34 45 50  & L  & 0.00    & 0.05 &   0.13  & 0.00   \nl
 16 15 15.0 &  47 03 52  & L  & 0.00    & 0.00 &   0.22  & 0.20    \nl
 16 16 50.3 &  47 42 50  & L  & $<0.03$ & 0.06 &   0.16  & 0.45    \nl
 16 28 44.7 &  42 24 32  & L  & 0.00    & 0.00 &   0.41  & 0.57    \nl
 16 32 01.7 &  37 30 27  & S  & $<0.06$ & 0.06 &   0.18  & $<0.25$ \nl    
 16 32 43.9 &  37 47 59  & L  & $<0.02$ & 0.04 &   0.08  & 0.26   \nl
 16 34 53.8 &  70 38 18  & S  & $<0.07$ & 0.14 &   0.33  & $<0.40$ \nl       
 16 34 57.5 &  46 52 24  & S  & $<0.06$ & 0.11 &   0.31  & $<0.33$ \nl     
 16 35 49.2 &  42 33 27  & L  & 0.00    & 0.00 &   0.27  & 0.36  & NC     \nl   
 16 39 35.6 &  39 23 16  & L  & $<0.02$ & 0.04 &   0.29  & 0.38         \nl
 16 50 56.2 &  40 00 51  & L  & $<0.02$ & 0.05 &   0.35  & 0.29       \nl
 16 53 43.2 & -80 27 53  & S  & $<0.09$ & 0.12 &   0.36  & $<0.76$ \nl
 16 55 12.9 &  27 55 44  & L  & 0.00    & 0.09 &   0.28  & 1.12    \nl
 17 02 19.0 &  60 37 52  & L  & 0.00    & 0.00 &   0.16  & 0.20       \nl          
 17 04 04.9 &  60 40 50  & S  & $<0.08$ & 0.15 &   0.49  &  0.78          \nl
 17 07 47.0 &  45 51 29  & L  & 0.02    & 0.20 &   0.32  & 27.11     \nl
 17 08 02.3 &  13 47 50  & L  & 0.08    & 0.32 &   1.68  &  2.03   \nl
 17 08 06.6 &  46 43 22  & L  & 0.00    & 0.06 &   0.18  & 0.00    \nl
 17 12 24.0 &  09 48 07  & L  & 0.08    & 0.09 &   0.16  & $<0.17$ \nl    
 17 19 10.3 &  50 25 23  & S  & $<0.08$ & 0.11 &   0.33  & $<0.38$ \nl     
 17 22 19.4 &  19 06 46  & L  & 0.00    & 0.11 &   0.62  & 0.52     \nl
 17 23 35.2 &  46 02 38  & L  & $<0.04$ & 0.14 &   1.22  & 1.42  & NC    \nl
 17 28 51.0 &  07 44 04  & L  & 0.00    & 0.09 &   0.13  & 0.00        \nl        
 17 31 28.8 &  59 58 58  & S  & $<0.08$ & 0.19 &   0.46  &  0.91         \nl
 17 32 03.7 &  13 11 40  & S  & $<0.05$ & 0.29 &   0.18  & $<0.30$ \nl                
 17 43 34.1 &  31 41 24  & L  & $<0.03$ & 0.08 &   1.73  & 2.60     \nl
 17 46 46.3 &  56 00 06  & L  & 0.00    & 0.00 &   0.26  & 0.39        \nl        
 17 47 28.0 &  56 12 00  & L  & $<0.03$ & 0.12 &   0.76  & 0.78           \nl     
 17 49 21.1 &  24 57 55  & L  & 0.00    & 0.00 &   0.40  & 0.48        \nl
 17 54 00.2 &  18 59 40  & L  & 0.26    & 0.08 &   0.11  & $<0.20$ \nl                
 17 54 22.1 &  42 01 30  & S  & $<0.05$ & 0.14 &   0.11  & $<0.30$ \nl          
 17 54 26.5 &  43 47 14  & L  & 0.00    & 0.09 &   0.40  & 0.40 & NC           \nl      
 17 54 56.5 &  70 58 42  & L  & 0.00    & 0.06 &   0.13  & 0.00    \nl
 17 55 41.2 &  43 29 47  & L  & 0.25    & 0.07 &   0.07  & 0.30       \nl          
 17 56 06.1 &  67 20 29  & L  & 0.00    & 0.08 &   0.56  & 0.79       \nl
 18 11 12.8 &  53 39 44  & L  & 0.00    & 0.00 &   0.48  & 0.63  & NC    \nl
 18 11 21.7 &  53 34 06  & L  & 0.00    & 0.00 &   0.83  & 1.00        \nl        
 18 24 56.7 &  50 54 16  & L  & 10.82   & 4.36 &   0.75  & 0.65           \nl       
 18 26 45.4 &  50 44 21  & L  & $<0.04$ & 0.08 &   0.98  & 1.30  & NC    \nl
 18 33 28.4 & -65 28 10  & S  &  0.54   & 2.24 &   2.19  & 1.40 \nl
 18 35 48.3 &  43 31 59  & S  & $<0.05$ & 0.05 &   0.14  & $<0.34$ \nl     
 18 38 30.5 &  42 53 18  & L  & 0.00    & 0.00 &   0.15  & 0.16 & NC   \nl              
 18 39 47.5 &  42 56 46  & L  & 0.00    & 0.13 &   0.35  & 0.00  & NC    \nl
 18 45 03.0 &  59 07 46  & L  & 0.00    & 0.00 &   0.64  & 0.83   \nl
 18 54 24.7 & -55 20 30  & L  & $<0.02$ & 0.11 &   0.33  & 2.63   \nl
 19 02 39.0 & -62 09 37  & L  & 0.00    & 0.00 &   0.23  & 0.33  \nl
 19 05 06.7 & -62 33 33  & L  & $<0.03$ & 0.13 &   0.39  & 0.72     \nl   
 19 24 28.0 & -41 40 42  & L  & 0.00    & 0.40 &   1.80  & 1.73  & NC       \nl         
 19 44 59.0 &  69 59 56  & L  & $<0.02$ & 0.15 &   0.26  & 0.82  & NC    \nl
 19 46 32.5 & -67 29 43  & S  & $<0.08$ & 0.16 &   0.49  & $<0.56$ \nl
 19 47 59.7 & -68 21 26  & L  & $<0.03$ & 0.06 &   0.58  & 0.823   \nl
 20 04 33.6 & -66 38 21  & L  & 0.54    & 0.17 &   0.16  & $<0.14$ \nl   
 20 11 07.2 & -61 03 47  & L  & 0.00    & 0.15 &   0.27  & 0.00      \nl           
 20 11 38.3 & -71 44 27  & L  & 0.00    & 0.00 &   0.74  & 0.96 & NC   \nl
 20 14 16.8 & -12 15 26  & L  & $<0.08$ & 0.35 &   1.18  & 5.87    \nl
 20 28 59.4 & -12 53 12  & L  & 1.07    & 1.19 &   0.83  & $<0.22$ \nl   
 20 46 54.4 & -52 42 17  & L  & 0.00    & 0.00 &   0.16  & 0.13   \nl
 21 13 24.8 & -39 22 32  & L  & 0.00    & 0.00 &   0.11  & 0.14   \nl
 21 27 32.8 & -43 21 58  & L  & $<0.03$ & 0.08 &   1.00  & 0.92   \nl
 21 33 01.1 & -64 51 09  & L  & 0.00    & 0.09 &   0.26  & 0.00  & NC     \nl  
 21 33 30.5 &  18 16 33  & L  & 0.00    & 0.00 &   0.16  & 0.21         \nl        
 21 47 40.9 & -27 06 13  & L  & $<0.04$ & 0.16 &   1.18  & 1.62  & NC    \nl
 21 54 12.6 & -07 59 43  & L  & 0.00    & 0.17 &   0.67  & 0.56     \nl
 21 56 01.6 & -31 04 41  & L  & 0.08    & 0.11 &   0.21  & 0.50  & NC   \nl
 21 59 26.8 & -51 07 49  & L  & 0.00    & 0.00 &   0.20  & 0.25  & NC      \nl           
 22 03 01.8 & -71 46 22  & L  & 0.00    & 0.00 &   0.18  & 0.20   \nl
 22 09 25.2 &  15 47 30  & L  & 1.37    & 0.37 &   0.27  & $<0.21$ \nl                
 22 21 14.5 & -02 21 34  & S  & $<0.09$ & 0.33 &   0.27  & 0.59    \nl
 22 23 11.7 & -05 12 04  & S  & $<0.11$ & 0.37 &   0.88  & 1.66    \nl
 22 51 08.1 & -18 05 20  & L  & 0.15    & 0.20 &   0.15  & $<0.18$ & NC  \nl   
 22 55 42.9 &  07 33 10  & L  & $<0.04$ & 0.15 &   0.52  & 1.06      \nl   
 23 03 41.4 &  11 09 14  & L  & $<0.03$ & 0.12 &   0.80  & 1.12      \nl
 23 12 50.4 & -59 19 38  & L  & 0.24    & 1.60 &  11.98  & 14.02        \nl         
 23 50 24.3 &  26 12 33  & L  & 0.00    & 0.00 &   0.35  & 0.42        \nl
 23 51 03.2 & -68 27 08  & L  & 0.00    & 0.00 &   0.32  & 0.35   \nl
 23 59 14.2 & -77 13 46  & L  & 0.00    & 0.08 &   0.27  & 0.60  \nl
\tablecomments{Fluxes are in Jy. The notation NC shows objects for which
confirmation through multiple scans was not possible.}
\enddata
\end{deluxetable}

\makeatletter
\def\jnl@aj{AJ}
\ifx\revtex@jnl\jnl@aj\let\tablebreak=\nl\fi
\makeatother
\begin{deluxetable}{cccccccccc}
\footnotesize
\tablecaption{Spectroscopic Identifications}
\tablehead{
\colhead{Object}           & \colhead{$\alpha_{2000}$} & 
\colhead{$\delta_{2000}$} & \colhead{$z$}      &
\colhead{Instrument}          & \colhead{Class}  &
\colhead{Comments} }        
\startdata
00168+2637 &  00 19 27.5   & +26 54 24 & 0.06337 & WHT &  H II \nl
00179+1503 &  00 20 32.6   & +15 20 51 & 0.03953 & WHT &  H II  & reddened \nl
00556+2937 &  00 58 21.4   & +29 53 28 & 0.08070 & KP2m & H II \nl 
01165+3146 &  01 19 22.0   & +32 02 39 & 0.05743 & WHT &  Sy1 \nl 
02033+3205 &  02 06 16.4   & +32 19 09 & 0.03219 & KP2m & LINER \nl 
02052+0228 &  02 07 49.9   & +02 42 55 & 0.15569 & KP2m & Sy1/QSO \nl 
05254-4010 &  05 27 08.8   & -40 07 24 & 0.17447 & EFOSC & H II & old population \nl
05261-2040 &  05 28 14.0   & -20 37 48 & 0.02851 & EFOSC & LINER \nl
06229-6434 &  06 23 07.6   & -64 36 20 & 0.12888 & EFOSC & Sy1 & strong Fe II \nl
06361-6217 &  06 36 35.8   & -62 20 33 & 0.16005 & EFOSC & LINER \nl
08020+1055 &  08 04 46.4   & +10 46 37 & 0.03549 & EFOSC & Sy1 & reddened \nl
08339+6517 &  08 38 22.9   & +65 07 16 & 0.0187  & INT   &      & \nl 
09426-1929 &  09 44 59.7   & -19 42 45 & 0.0327  & INT   & starburst & Arp 252, pair \nl
10149+2102 &  10 17 40.7   & +20 47 44 & 0.10913 & KP4m  & H$\alpha$ only \nl
10360-0654 &  10 38 33.9   & -07 10 14 & 0.00729 & EFOSC & starburst & strong WR lines \nl
10570+7239 &  11 00 38.4   & +72 22 51 & 0.02889 & KP4m  & Sy1 & reddened \nl
11507+5823 &  11 53 26.4   & +58 06 45 & 0.06520 & KP4m  & Sy2 \nl   
11549+7327 &  11 57 35.0   & +73 10 38 & 0.51100 & KP4m  & QSO & strong Fe II \nl  
12062+6811 &  12 08 42.0   & +67 54 48 & 0.23199 & WHT   & H$\alpha$ only \nl
12108+3703 &  12 13 16.5   & +36 47 34 & 0.08742 & KP4m  & Sy2 \nl
12179+3013 &  12 20 27.3   & +29 57 05 & 0.06065 & FOS   & Sy1 & reddened \nl
12215+1107 &  12 24 02.8   & +10 50 56 & 0.02568 & FOS   & H$\alpha$ only  \nl
12238+4846 &  12 26 15.7   & +48 29 39 & 0.00136 & KP4m  & starburst \nl
12266+3240 &  12 29 06.9   & +32 24 18 & 0.17179 & KP4m  & Sy2/LINER & stellar absorption \nl 
12295+1413 &  12 32 04.5   & +13 57 22 & 0.06345 & FOS   & Sy2 \nl 
12397+3333 &  12 42 10.5   & +33 17 03 & 0.04436 & FOS   & Sy2 \nl 
12410+3923 &  12 43 25.9   & +39 07 23 & 0.09990 & WHT   & H II & reddened \nl
12498+2721 &  12 52 17.7   & +27 05 08 & 0.02128 & KP4m  & LINER \nl
12583+2847 &  13 00 40.7   & +28 31 12 & 0.03037 & KP4m  & starburst \nl 
13207+5510 &  13 22 49.2   & +54 55 29 & 0.06604 & KP4m  & Sy1 \nl 
13556+6951 &  13 56 47.4   & +69 37 12 & 0.03496 & IPCS  & Sy2 \nl 
14030+5338 &  14 04 52.6   & +53 23 33 & 0.08103 & KP4m  & Sy 2 \nl
14071+5446 &  14 08 55.5   & +54 43 05 & 0.04116 & KP4m  & Sy2/H II   & reddened \nl
14135+1334 &  14 15 58.8   & +13 20 24 & 0.2467  & WHT   &           \nl
14390+3524 &  14 41 05.8   & +35 11 40 & 0.12501 & KP4m  & H II \nl 
14399+3605 &  14 42 01.0   & +35 52 58 & 0.14787 & WHT   & H II \nl
14400+3539 &  14 42 07.4   & +35 26 23 & 0.07755 & KP4m  & Sy 1 & strong Fe II \nl 
15033+2715 &  15 05 34.9   & +27 03 30 & 0.26156 & KP4m  & H II \nl
15112+1108 &  15 13 42.4   & +10 56 53 & 0.06635 & KP4m  & Sy2 \nl  
15126+3640 &  15 14 37.5   & +36 29 39 & 0.06426 & WHT   & H$\alpha$ only \nl
15216+3015 &  15 23 42.1   & +30 05 02 & 0.11041 & WHT   & Sy 2 \nl 
15261+5502 &  15 27 26.8   & +54 51 51 & 0.22803 & WHT   & H II \nl  
15347+5804 &  15 35 52.3   & +57 54 11 & 0.03077 & KP4m  & Sy1 \nl 
15357+5934 &  15 36 48.0   & +59 23 52 & 0.01017 & KP4m  &      & NGC 5976; reddened \nl
15561+2600 &  15 58 18.7   & +25 51 25 & 0.07218 & FOS   & Sy1 \nl 
15569+2640 &  15 59 05.6   & +25 32 12 & 0.07169 & KP4m  & H II & reddened \nl
16118+3445 &  16 13 42.2   & +34 38 33 & 0.26824 & WHT   & Sy2 \nl  
16168+4742 &  16 18 18.3   & +47 35 31 & 0.10967 & IPCS  & H II \nl 
16287+4224 &  16 30 19.3   & +42 18 21 & 0.07000 & WHT   & H II \nl 
16320+3730 &  16 33 45.0   & +37 23 36 & 0.17490 & KP4m  & Sy2 \nl  
16327+3747 &  16 34 30.8   & +37 41 44 & 0.09954 & KP4m  & Sy2 \nl 
16349+4652 &  16 36 20.4   & +46 46 39 & 0.22315 & WHT   & H$\alpha$ only \nl
16395+3923 &  16 41 15.1   & +39 17 26 & 0.03134 & KP4m  & H II \nl
16509+4000 &  16 52 34.7   & +39 55 46 & 0.27616 & WHT   & H II \nl 
16552+2755 &  16 57 16.2   & +27 50 59 & 0.03366 & KP4m  & Sy   & NGC 6264 (W of 2) \nl  
17023+6037 &  17 02 56.7   & +60 33 47 & 0.1041  & KP4m  & H$\alpha$/[N II] only \nl
17040+6040 &  17 04 39.0   & +60 36 51 & 0.16538 & WHT   & H$\alpha$ only \nl
17077+4551 &  17 09 12.6   & +45 47 45 & 0.06932 & KP4m  & H$\alpha$/[N II] only \nl
17080+1347 &  17 10 18.2   & +13 44 06 & 0.0312  & KP4m  & H II \nl
17081+4643 &  17 09 28.3   & +46 39 35 & 0.31516 & WHT   & Sy2 \nl   
17191+5025 &  17 20 29.5   & +50 22 37 & 0.02441 & KP4m  & Sy2 \nl 
17223+1906 &  17 24 33.2   & +19 04 10 & 0.05398 & KP4m  & H II \nl
17235+4602 &  17 25 00.4   & +45 59 46 & 0.06290 & WHT   & H II \nl
17314+5958 &  17 32 12.4   & +59 56 27 & 0.02747 & KP4m  & Sy & Akn 522 \nl 
17435+3141 &  17 45 28.3   & +31 40 21 & 0.07349 & WHT   &    & reddened; comp at matching $z$ \nl
17467+5600 &  17 47 35.6   & +55 58 52 & 0.11273 & KP4m  & H$\alpha$/[N II] only \nl
17493+2457 &  17 51 24.1   & +24 57 18 & 0.07259 & WHT   & H II \nl
17543+4201 &  17 55 56.9   & +42 00 59 & 0.22888 & WHT   & LINER? \nl
17544+4347 &  17 55 58.1   & +43 47 12 & 0.07288 & KP4m  & Sy 2 \nl
17561+6720 &  17 55 55.2   & +67 20 05 & 0.05079 & KP4m  &           & reddened \nl
18112+5339 &  18 12 16.7   & +53 40 37 & 0.17016 & WHT   & LINER \nl  
18397+4256 &  18 41 18.5   & +42 59 42 & 0.04967 & KP4m  &        \nl   
19051-6233 &  19 09 38.9   & -62 28 54 & 0.08538 & EFOSC &        & ESO 104-IG41 \nl 
20142-1215 &  20 17 06.3   & -12 05 51 & 0.01995 & WHT   & LINER \nl 
21330-6451 &  21 37 00.1   & -64 37 59 & 0.07091 & EFOSC & QSO/Sy 1 & strong Fe II \nl
21542-0759 &  21 56 49.5   & -07 45 33 & 0.05475 & WHT   & H II \nl 
22212-0221 &  22 23 49.5   & -02 06 12 & 0.05566 & WHT   & Sy 1 & double radio source \nl
22557+0733 &  22 58 17.4   & +07 48 54 & 0.02876 & KP2m  & H II \nl
23036+1109 &  23 06 12.5   & +11 25 37 & 0.05745 & KP2m  & H II \nl 
23504+2612 &  23 52 59.7   & +26 29 41 & 0.02741 & KP2m  & H II \nl 
\enddata
\end{deluxetable}

\makeatletter
\def\jnl@aj{AJ}
\ifx\revtex@jnl\jnl@aj\let\tablebreak=\nl\fi
\makeatother
\begin{deluxetable}{lclclclc}
\tablecaption{$R$-Band Photometry of Program Sources}
\tablehead{
\colhead{Object}           & \colhead{$R$}      &
\colhead{Object}           & \colhead{$R$}      &
\colhead{Object}           & \colhead{$R$}      &
\colhead{Object}           & \colhead{$R$} }
\startdata       
00168+2637 & 15.70 & 11507+5823 & 15.77 & 15112+1108  & 16.37 & 17235+4602 & 15.46  \nl
00179+1503 & 14.94 & 11549+7327 & 16.27 & 15126+3640 & 15.08 & 17314+5958 & 13.40   \nl
00556+2937 & 16.30 & 12062+6811 & 20.43 & 15216+3015 & 16.27 & 17435+3141 & 16.62  \nl
01165+3146 & 14.67 & 12108+3703 & 15.45 & 15261+5502 & 18.22 & 17493+2457 & 17.20  \nl
01195+8649 & 16.61 & 12179+3013 & 16.53 & 15347+5804 & 14.56 & 17543+4201 & 18.77  \nl
01501-2334 & 14.94 & 12215+1107 & 15.35 & 15357+5934 & 14.04 & 17561+6720 & 15.53  \nl
02033+3205 & 15.44 & 12266+3240 & 16.67 & 15561+2600 & 15.54 15.49 & 18112+5339 & 18.23  \nl
02052+0228 & 15.03 & 12295+1413  & 16.06 & 15569+2540 & 15.49 & 18267+5044 & 17.85  \nl
03062+1051 & 14.63 & 12397+3333 & 14.46 & 16118+3445 & 17.61 & 18397+4256 & 15.28  \nl
03533+2606 & 17.05 & 12410+3923 & 17.31 & 16152+4703 & 20.52 & 20142-1215 & 12.77  \nl
0353+261 & 19.04 & 12498+2721 & 14.34 & 16168+4742 & 17.63 & 21476-2706 & 16.40  \nl
05254-4010 & 17.79 & 12583+2847 & 15.56 & 16287+4224 & 17.38 & 22212-0221 & 14.67  \nl
05261-2040 &  16.36 & 13207+5510 & 14.34 & 16320+3730 & 16.99 & 22231-0512 & 14.48  \nl
06229-6434 &  15.05 & 13446-3625 & 16.02 & 16327+3747 & 17.76 & 22557+0733 & 15.35  \nl
08020+1055 & 13.09 & 13556+6951 & 14.90 & 16395+3923 & 15.41 & 23036+1109 & 15.98  \nl
08228+2927 & 18.26 & 14030+5338 & 15.68 & 16509+4000 & 19.09 & 23504+2612 & 16.46  \nl
08339+6517 &  12.64 & 14071+5446 & 14.02 & 16552+2755 & 13.76 \nl
08370+2959 & 14.21 & 14390+3524 & 16.55 & 17040+6040 & 19.19  \nl
09426-1929 & 14.56 & 14399+3605 & 18.66 & 17081+4643 & 19.51  \nl
10360-0654 &  13.05 & 14400+3539 & 13.91 & 17191+5025 & 13.72  \nl
10570+7239 & 16.05 & 15033+2715 & 19.93 & 17223+1906 & 15.46  \nl

\enddata
\end{deluxetable}

\makeatletter
\def\jnl@aj{AJ}
\ifx\revtex@jnl\jnl@aj\let\tablebreak=\nl\fi
\makeatother
\begin{deluxetable}{lccrl}
\tablecaption{Radio Identifications of Warm IRAS-AO Sources}
\tablehead{
\colhead{Object}           & \colhead{$\alpha_{2000}$}      &
\colhead{$\delta_{2000}$}           & \colhead{Flux (mJy)}      &
\colhead{Wavelength (cm)} }
\startdata
01215-3519 & 01 23 54.57  & -35 03 57 &  13.1 & 20 NVSS\nl
01501-2334 & 01 52 27.00  & -23 19 52 &   4.7 & 20 NVSS\nl
02317-3905 & 02 33 44.65  & -38 52 21 & 2.9 & 20 NVSS\nl
03062+1051 & 03 08 56.70  & +11 03 16 &  5.38 & 6 \nl
%                                       12.1  & 20 NVSS
05116-3323 & 05 13 24.71  & -33 20 07 &   4.8 & 20 NVSS\nl
05239-0626 & 05 26 19.02  & -06 24 04 &   4.7 & 20 NVSS\nl
05254-4010 & 05 27 08.95  & -40 07 18 &   5.5 & 20 NVSS\nl
05563-3820 & 05 58  2.20  & -38 20 02 &  34.6 & 20 NVSS\nl
06555+5415 & 06 59 38.14  & +54 11 47  &  16.1 & 20 FIRST\nl
08020+1055 & 08 04 46.40  & +10 46 36  &  21.9 & 20 FIRST\nl
%08226+2927 & 08 25 55.07 & +29 17 31.0 &  & 20 FIRST\nl
08339+6517 & 8 38 23.29   & +65 07 14  & 33.7 & 20 NVSS\nl
%08370+2959 & 08 40 02.28 & +29 49 00.5 &  364 & 20 FIRST triple\nl %NED radio gal
%08519+2018 & 08 54 48.88 & +20 06 30.5 &  1512 & 20 FIRST\nl %NED BL Lac object
09328-1436 & 09 35 11.36  & -14 50 02  &     9.1 & 20 NVSS\nl
09426-1929 & 09 44 59.76  & -19 42 44  &    97.6 & 20 NVSS\nl
09442-0744 & 09 46 45.88  & -07 59 44  &     3.4 & 20 NVSS\nl
10360-0654 & 10 38 33.62  & -07 10 14  & 51.5 & 20 FIRST\nl
%11507+5823 & 11 53 26.49  & +58 06 44  &  8.5 & 20 FIRST\nl
11507+5823 & 11 53 04.06  & +57 50 02  &  2.90 & 6 \nl
11549+7327 & 11 57 35.54  & +73 10 36  & 21.28 & 6 \nl
%       "                                 13.6  & 20 NVSS \nl
12062+6811 & 12 08 42.28  & +67 54 56  &  2.3  & 20 NVSS \nl
12108+3703 & 12 13 16.50  & +36 47 33 &  1.64 & 6 \nl
12295+1413 & 12 32 04.50  & +13 57 22  &  1.13 & 6 \nl
%12108+3703 & 12 13 16.61 & +36 47 34.5 &  & 20 FIRST\nl   %use 6cm
12215+1107 & 12 24 02.89 & +10 50 56  &  1.8 & 20 FIRST\nl
12238+4646 & 12 26 17.89 & +48 29 26 &  4.5 & 20 FIRST\nl
12266+3240 & 12 29 06.90 & +32 24 18 &  3.0 & 20 FIRST\nl
12295+1413 & 12 32 04.50 & +13 57 22 &  1.9 & 20 FIRST\nl
12397+3333 & 12 42 10.63 & +33 17 03 &  6.3 & 20 FIRST\nl
%12498+2721 & 12 52 13.92 & +27 03 46.2 &  & 20 FIRST\nl
12583+2847 & 13 00 40.77  & +28 31 16 &  6.3 & 20 NVSS\nl
13207+5510 & 13 22 49.21 & +54 55 28 &  8.9 & 20 FIRST\nl
13556+6951 & 13 56 40.77  & +69 35 42 & 28.0 & 20 NVSS \nl
14030+5338 & 14 04 52.63 & +53 23 32 &  1.9 & 20 FIRST\nl
%14135+1334 & 14 15 58.82 & +13 20 23.8 &  1179 & 20 FIRST\nl% PKS QSO
1439+354 & 14 41 08.41 & +35 11 52  &  0.74 & 6 \nl
         & 14 41 05.82 & +35 11 40 &  0.64 & 6 \nl
         & 14 41 05.50 & +35 10 08 &  0.97 & 6 \nl
%14399+3605 & 14 41 55.22 & +35 52 43 &  & 20 FIRST\nl %????
14400+3539 & 14 42 07.47 & +35 26 23 &  3.3 & 20 FIRST\nl
15033+2715 & 15 05 30.30 & +27 04 10 &  2.1 & 20 FIRST\nl
15112+1108 & 15 13 42.37 & +10 56 53 &  7.6 & 20 FIRST\nl
15126+3640 & 15 14 33.12 & +36 29 42 &  0.8 & 20 FIRST\nl
%15216+3015 & 15 23 42.21 & +30 05 02 &  2.7 & 20 FIRST\nl
15216+3015 & 15 23:42.09 & +30 05 02 &  0.58 & 6 \nl
15261+5502 & 15 27 26.70 & +54 51 51 &  1.5 & 20 FIRST\nl
15347+5804 & 15 35 52.42 & +57 54 10 &  5.1 & 20 FIRST\nl
15562+2600 & 15 58 18.75 & +25 51 25 &  2.4 & 20 FIRST\nl
15569+2540 & 15 59 05.63 & +25 32 12 &  2.1 & 20 FIRST\\nl
16075+2958 & 16 09 29.47 & +29 50 51 &  0.6 & 20 FIRST\nl
%16118+3445 & 16 13 42.25 & +34 38 32 &  3.4 & 20 FIRST\nl
16118+3445 & 16 13 42.24 & +34 38 32 &  1.11 & 6 \nl
16152+4703 & 16 16 42.65 & +46 56 48 &  4.1 & 20 FIRST\nl
16287+4224 & 16 30 19.38 & +42 18 19 &  2.0 & 20 FIRST\nl%
16320+3730 & 16 33 44.98 & +37 23 35 &  1.0 & 20 FIRST\nl
16320+3730 & 16 33 45.31 & +37 22 53 &  0.42 & 6 \nl
16349+4652 & 16 36 18.11 & +46 46 30 &  2.2 & 20 FIRST\nl
16552+2755 & 16 57 10.08 & +27 50 14 &  2.0 & 20 FIRST\nl
%17077+4551 & 17 09 12.74 & +45 47 44 &  1.1 & 20 FIRST\nl
17077+4551 & 17 09 12.59 & +45 47 44 &  0.46 & 6 \nl
%17081+4643 & 17 09 28.37 & +46 39 34 & 11.6 & 20 FIRST\nl
17081+4643 & 17 09 28.32 & +46 49 35 &  5.76 & 6 \nl
17124+0948 & 17 14 43.17 & +09 45 00 &  0.93 & 6 \nl
%                                         2.3 & 20 NVSS\nl
17191+5025 & 17 20 29.57 & +50 22 04 &  5.2 & 20 FIRST\nl
17223+1906 & 17 24 33.33  & +19 04 01 &  4.0 & 20 NVSS\nl
17235+4602 & 17 25 00.37  & +45 59 45 &  4.5 & 20 FIRST\nl
17435+3141 & 17 45 28.39  & +31 40 19 &   7.4 & 20 NVSS\nl
17549+7058 & 17 56 05.97 & +42 00 59 &  0.44 & 6 \nl
18112+5339 & 18 12 17.22  &  +53 40 29  &   2.4 & 20 NVSS\nl
18267+5044 & 18 28 00.28  & +50 46 20 &  32.5 & 20 NVSS\nl
18358+4331 & 18 37 26.49 & +43 34 52 &  0.43 & 6 \nl
18397+4256 & 18 41 18.50 & +42 59 41 &  1.86 & 6 \nl
19449+6959 & 19 45 04.77 & +70 06 39 &  0.71 & 6 \nl
           & 19 44 47.02 & +70 06 54 &  0.41 & 6 \nl
20142-1215 & 20 17  6.44  & -12 05 51   &  28.2 & 20 NVSS\nl
20289-1253 & 20 31 49.22  & -12 42 38 &  1.46 & 6 \nl
21476-2706 & 21 50 34.42  & -26 52 10 &   5.2 & 20 NVSS\nl
21542-0759 & 21 56 49.52  & -07 45 32 &  3.7 & 20 FIRST\nl
21560-3104 & 21 58 54.97  & -30 50 46  &   3.9 & 20 NVSS \nl
22511-1805 & 22 53 56.60 & -17 49 14 &  1.24 & 6 \nl   
%22212-0221 & 22 23 49.54 & -02 06 13 & 2059 & 20 FIRST; 3C 445\nl
23036+1109 & 23 06 12.46  & +11 25 41  &    5.8 & 20 NVSS\nl
\enddata
\end{deluxetable}

\makeatletter
\def\jnl@aj{AJ}
\ifx\revtex@jnl\jnl@aj\let\tablebreak=\nl\fi
\makeatother

\begin{deluxetable}{lccclcccc}
\small
\tablecaption{Stellar Identifications of Warm IRAS-AO Sources}
\tablehead{
\colhead{Object}           & \colhead{$\alpha_{1950}$}      &
\colhead{$\delta_{1950}$}           & \colhead{ID} &
\colhead{Spectral Type}   & \colhead{$V$} & \colhead{$B-V$}
& \colhead{Distance, pc} }
\startdata
02204+3512 & 02 20 26.2 & +35 12 38 & HD 14735 &  G5 &  6.76 & 1.16 & 225  \nl            
05239-0626 & 05 23 56.2 & -06 26 30 & BD -6$^\circ$ 1178 & & 10.41 & 0.38 \nl
05356-2839 & 05 35 40.8 & -28 39 10 & HD 37484 &  F3V & 7.25 & 0.38 & 60 \nl
10119-0052 & 10 11 59.6 & -00 52 36 & HD 88803 & K2  & 8.12 & 1.18 & 255 \nl
16075+2958 & 16 07 32.9 & +29 58 49 & HD 145373 & K5  & 8.33 & 1.47 \nl
17320+1311 & 17 32 03.7 & +13 11 40 & 54 Oph  &  G8III & 6.54 & 0.94 & 156 \nl
17540+1859 & 17 54  0.2 & +18 59 40 & BD +19$^\circ$ 3472 & K2 & 9.34 & 1.53 \nl
17556+4329 & 17 55 41.2 & +43 29 47 & BD +43$^\circ$ 2847 & K5  & 9.07 & 1.52 \nl
20045-6638 & 20 04 33.6 & -66 38 21 &                     &      &     &   & \nl
22094+1547 & 22 09 25.2 & +15 47 30 & HD 210702 & K1 III & 5.94 & 0.96 \nl

\enddata
\end{deluxetable}

\makeatletter
\def\jnl@aj{AJ}
\ifx\revtex@jnl\jnl@aj\let\tablebreak=\nl\fi
\makeatother
\begin{deluxetable}{c c c l l}
\tablecaption{60$\mu$m Sources in the Hercules Deep Observation}
\tablehead{
\colhead{Number} & 
\colhead{$\alpha_{1950}$}      &
\colhead{$\delta_{1950}$}           & \colhead{F(60$\mu$m, Jy)}      &
\colhead{Identification} }
\startdata
1  & 16 32 26.1 & +37 26 23  & 0.07  & disk galaxy $z=0.0292$; FIRST source 51" E \nl
2  & 16 33 26.4 & +37 26 47  & 0.68  & Mira variable W Herculis \nl
3  & 16 34 02.1 & +37 28 50  & 0.055 \nl
4  & 16 32 00.3 & +37 29 52  & 0.18   & IRAS-S 16320+3730 (Sy2) $z=0.1749$\nl
5  & 16 33 58.0 & +37 39 35  & 0.055 \nl
6  & 16 33 23.7 & +37 40 32  & 0.013 \nl
7  & 16 33 55.8 & +37 42 57  & 0.022 & very faint; FIRST ID \nl
8  & 16 32 45.1 & +37 47 48  & 0.063  & IRAS-L 16327+3747 (Sy2) $z=0.0995$\nl
9  & 16 32 40.5 & +37 53 13  & 0.021  & starburst galaxy $z=0.0974$ \nl
10 & 16 31 56.7 & +37 53 36  & 0.020 \nl
11 & 16 34 16.1 & +37 52 43  & 1.129  & 11 mJy FIRST source at 77" \nl
12 & 16 33 09.6 & +37 55 39  & 0.14 \nl
13 & 16 33 50.4 & +37 57 06  & 0.044 \nl
14 & 16 32 03.6 & +38 01 09  & 0.045 \nl
15 & 16 33 36.5 & +38 02 31  & 0.012  & FIRST: bright outlier to group/cluster \nl
16 & 16 34 17.1 & +38 05 13  & 0.039 \nl
17 & 16 32 22.2 & +38 05 50  & 0.046 \nl
18 & 16 32 41.3 & +38 10 38  & 0.10 \nl
19 & 16 32 05.1 & +38 14 10  & 0.064 \nl 
20 & 16 33 31.4 & +38 14 30  & 0.014 & blazar B2 1633+28, 4C +38.41 $z=1.814$\nl
21 & 16 32 13.3 & +38 19 48  & 0.025 \nl
22 & 16 32 50.7 & +38 27 02  & 0.040 \nl
23 & 16 33 40.2 & +38 30 20  & 0.031 \nl
24 & 16 32 04.6 & +38 33 26  & 0.120 \nl
25 & 16 34 13.7 & +38 39 53  & 0.092 \nl
26 & 16 32 28.6 & +38 41 58  & 0.059 & faint, diffuse; FIRST ID \nl
27 & 16 33 00.5 & +38 43 56  & 0.191 \nl
28 & 16 33 36.0 & +38 51 12  & 0.044 \nl
29 & 16 34 20.4 & +38 52 32  & 0.041 \nl
30 & 16 32 23.0 & +39 00 02  & 0.088 \nl
31 & 16 34 05.3 & +39 03 36  & 0.022 \nl

\enddata
\end{deluxetable}

\end{document}